\documentclass[11pt, conference, compsocconf]{IEEEtran}
\IEEEoverridecommandlockouts
\usepackage{cite}
\usepackage{graphicx}
\usepackage{doi}
\usepackage{amsmath}
\usepackage{xcolor}
\usepackage[standard]{ntheorem}
 
\usepackage{graphicx}
\usepackage{doi}
\usepackage{comment}
\usepackage[ruled,vlined]{algorithm2e}
\usepackage{xcolor}

\DeclareMathOperator{\tr}{tr}
\DeclareMathOperator{\kappaa}{\Psi}

\DeclareMathOperator{\U}{\mathbb{U}}
\DeclareMathOperator{\I}{\mathbb{I}}
\DeclareMathOperator{\G}{\mathbb{G}}
\DeclareMathOperator{\V}{\mathbb{V}}
\DeclareMathOperator{\C}{\mathbb{C}}

\DeclareMathOperator{\PP}{\mathbb{P}}

\DeclareMathOperator{\X}{\mathbb{X}}

\DeclareMathOperator{\B}{\mathbb{B}}

\begin{document}

\title{A quantum learning approach based on Hidden Markov Models for failure scenarios generation}

\author{\IEEEauthorblockN{Ahmed ZAIOU}
\IEEEauthorblockA{\textit{EDF Lab Saclay, PERICLES, France }\\
\textit{LIPN - CNRS UMR 7030, USPN} \\
\textit{LaMSN, La Maison des Sciences Numériques}\\
\textit{ ahmed.zaiou@edf.fr }
}
\and
\IEEEauthorblockN{ Younès BENNANI}
\IEEEauthorblockA{\textit{LIPN - CNRS UMR 7030, Université Sorbonne Paris Nord} \\
\textit{LaMSN, La Maison des Sciences Numériques}\\
younes.bennani@sorbonne-paris-nord.fr}
\and

\IEEEauthorblockN{ Basarab MATEI}
\IEEEauthorblockA{\textit{LIPN - CNRS UMR 7030, Université Sorbonne Paris Nord} \\
\textit{LaMSN, La Maison des Sciences Numériques, France}\\
basarab.matei@sorbonne-paris-nord.fr}

\and

\IEEEauthorblockN{ Mohamed HIBTI}
\IEEEauthorblockA{\textit{EDF Lab Saclay, PERICLES} \\
France\\
mohamed.hibti@edf.fr}

}

\maketitle

\begin{abstract}
Finding the failure scenarios of a system is a very complex problem in the field of Probabilistic Safety Assessment (PSA). In order to solve this problem we will use the Hidden Quantum Markov Models (HQMMs) to create a generative model. Therefore, in this paper, we will study and compare the results of HQMMs and classical Hidden Markov Models HMM on a real datasets generated from real small systems in the field of PSA. As a quality metric we will use Description accuracy DA and we will show that the quantum approach gives better results compared with the classical approach, and we will give a strategy to identify the probable and no-probable failure scenarios of a system. 
\end{abstract}

\begin{IEEEkeywords}
 Quantum Machine Learning (QML); Quantum Hidden Markov Models (QHMMs); Hidden Markov Models (HMMs); Probabilistic Safety Assessment (PSA)
\end{IEEEkeywords}

\IEEEpeerreviewmaketitle

\section{Introduction}
 
\noindent Quantum Computing \cite{jaeger2007quantum} is becoming increasingly a solution for complex problems that are hard to solve with classical computers \cite{preskill2012quantum}. This computational ability has been demonstrated theoretically by several papers, which suggests very interesting algorithms such as Grover's algorithm \cite{grover1996fast}, Shor algorithm \cite{Shor1997}  and several others such as \cite{pednault2019leveraging,zhong2020quantum}. Machine Learning is one of the most usable domains of Quantum Computing, several papers can be found in this field, either in Universal Quantum computers \cite{10.1007/978-3-030-36718-3_47, martin2022quantum} or in Quantum Annealing computers \cite{9659997,sjolund2022graph}. Also, in Deep Learning 
we find several proposed solutions \cite{dalgaard2022predicting, casares2022qfold}.
One of the most studied models recently is the  Quantum Hidden Markov Models (QHMMs), the paper \cite{adhikary2020expressiveness} demonstrates that QHMMs are a special subclass of the general class of Observable Operator Models (OOMs) and also provides a learning algorithm for HQMMs using the constrained Gradient Descent method. It also demonstrates that this approach is faster and more suitable for larger models compared to previous learning algorithms. In addition, there are other learning algorithm of QHMMs like \cite{srinivasan2018learning}, this work demonstrates some theoretical advantages of QHMMs compared to HMMs and also that each HMM of finite dimension can be modeled by a QHMM of finite dimension.
\\In this paper, we will discuss a very complex problem in the field of Probabilistic Safety Assessment (PSA) using QHMMs. This problem particularly concerns the searching and modeling Failure Scenarios of a system from an initial state to one of the Severe Failure States. Here, we will consider these failure scenarios as sequences, which gives us a sequential dataset, and we will use QHMMs to learn models that can provide us more scenarios and also detect probable and no-probable scenarios.
\\The rest of this paper is organized as follows: in section \ref{sq:HMM}, we describe the classical Hidden Markov Models.
In section \ref{sq:HQMM}, we describe Hidden Quantum Markov Models. 
In section \ref{sq:OC}, we discuss our approach, we talk about the process of building a dataset from the system architecture and also how to learn QHMMs from the datasets that we have and how to use these models to decide if a scenario is probable or not.
The results obtained after testing in four datasets are presented in section  \ref{sq:results}. 
Finally the paper ends with a conclusion.

\section{Classical Hidden Markov Models}
\label{sq:HMM}

\noindent In this section, we will briefly recall  the definition of the  classical Hidden Markov Model (HMM). A  hidden  Markov model  is  a  tool  for  representing  probability  distributions  over sequences of observations. It is assumed that each observation $X_t$ was generated by some process whose state $S_t \in [1..K]$ is unknown (hence the name hidden).
 \begin{definition}[HMM]
 \label{def:HMM}
A HMM is made of five key elements:
\begin{enumerate}
\item \textbf{An alphabet} $\Sigma=\{o_1,\cdots,o_M\}$.  
\item \textbf{A set of index of states} $\boldsymbol{Q}=\{1,\cdots,K\}$
\item \textbf{A transition  probability matrix} $A= (a_{kk'}),$ $A  \in \mathbb{R^{K\times K}}, \; \forall k \sum_{k'} a_{kk'}=1$,
$a_{kk'}$ is the probability of transition from state $k$ to state $k'.$

\item \textbf{Emission probabilities} within each state: 
$e_{k}(x)=\PP(x|Q=k),\; 
\sum_{x \in \Sigma} e_{i}(k)=1.$
\item \textbf{Starting probabilities}: $\pi_{1},\cdots,\pi_{K},$   $\sum_k\pi_{k}=1.$	
\end{enumerate}	
Therefore a  HMM model is denoted $M=\{A,B,\pi\}$, where:
$A$ is the transition probability matrix, 
$B$ contains the emissions probability laws $e_{k}(x)$,
$\pi$ the starting probabilities.
\end{definition}
\subsection{The Forward-Backward algorithm}
\noindent Given a HMM, we can generate a sequence of length $n$ as follows:
\begin{enumerate}
	\item Start at state $Q_1$ according to $\pi_1$
	\item Emit observation $x_1$ according to $e_{Q_1}(x_1)$
	\item Go to state $Q_2$ according to $a_{Q_1,Q_2}$
	\item ... until emitting $x_n$
\end{enumerate}
More precisely, given the model $M$, we want an algorithm that can compute the following probabilities:
\begin{itemize}
	\item $\PP(X)$ the probability of $X$ 
	\item $\PP(x_i \cdots x_j)$ the probability of a substring of $X$  		
	\item $\PP(Q_i=k|X)$ the posterior probability that the $i^{th}$ state is $k$, given $X$
\end{itemize}
In order to  calculate $\PP(X)$, the probability of the whole sequence given the HMM, we sum all possible ways of generating $X$:
$$\PP(X)= \sum_{\boldsymbol{Q}} \PP(X,\boldsymbol{Q}) = \sum_{\boldsymbol{Q}} \PP(X|\boldsymbol{Q})\PP(\boldsymbol{Q}).$$
To avoid summing over an exponential number of paths $\boldsymbol{Q}$, we define the \textbf{forward probability}:
$f_{k}(i) = \PP(x_1 \cdots x_t,Q_t=k).$
This represents  the probability of observing the sequence $x_1 \cdots x_t$ and having the $t^{th}$ state being $k$.
Straightforward computations give:
\begin{equation}
\label{forward}
f_{k}(t) = \PP(x_1 \cdots x_t,Q_t=k)=
e_{k}(x_t) \sum_l f_l (t-1) a_{lk}    
\end{equation}
The Equations \eqref{forward} represents the key of the  so-called  "Forward Algorithm" since we can compute all the $f_k(t)$ using dynamic programming as follows: firstly we initialize $f_0(0)=1$
and $\forall k>0 \quad f_k(0)=0$, secondly we iterate $f_k(i)=e_{k}(x_t) \sum_l f_l (t-1) a_{lk}$
and we terminate by putting $\PP(X)= \sum_{k} f_k(N).$
\\In order  to compute $\PP(Q_t=k|X)$ the probability distribution on the $t^{th}$ position given $X$, we proceed as follows: since we have $\PP(Q_t=k|X)=\frac{\PP(Q_t=k,X)}{\PP(X)}$, we start by computing $\PP(Q_t=k,X)=\PP(x_1 \cdots x_t , Q_t=k, x_{t+1} \cdots x_N)$, we obtain: 
$$
\PP(Q_t=k,X)= f_k(t) b_k(t),
$$
where $b_k(t)$ are  the {\bf  backward probability }
$b_{k}(t) = \PP(x_{t+1} \cdots x_N | Q_t=k)$. It follows that:
\begin{equation}
\label{backward}
b_{k}(t) = 	\sum_l e_l(x_{t+1})a_{kl} b_l(t+1).
\end{equation}
The Equations \eqref{backward} represents the key of the  so-called "Backward Algorithm" since we can compute all the $b_k(t)$ using dynamic programming as follows: firstly we initialize
$\forall k \quad b_k(N)=0$, secondly we iterate $b_k(i)=\sum_l e_l(x_{t+1})a_{kl} b_l(t+1)$ and we terminate by putting $P(X)= \sum_{l} \pi_l e_l(x_1) b_l(1).$
\subsection{The Baum-Welch algorithm}
\noindent The Forward-Backward algorithm can be adapted into an EM like procedure to learn the model. This is known as the Baum-Welch algorithm.

Repeat until convergence:
\begin{itemize}
	\item Compute the probability of each state at each position using forward and backward probabilities.
  This gives the expected distribution of the observations for each state using Bayes Theorem.

	\item Compute the probability of each pair of states at each pair of consecutive positions $t$ and $t+1$ using forward(t) and backward(t+1). This gives the expected transition counts.
\end{itemize}

\section{Quantum Hidden Markov Models}
\label{sq:HQMM} 

\noindent In this section we define the quantum version of HMM,  the Quantum Hidden Markov Model (QHMM). To 
represents  an  QHMM we  use  quantum circuits.

\begin{definition}[Quantum Hidden Markov Model]
A $K$-dimensional Quantum Hidden Markov Model $(K-HQMM)$ is made by :

\begin{enumerate}

\item \textbf{An alphabet}, a set of discrete observations $\Sigma=\{o_1,\cdots,o_M\}$.
\item \textbf{A set of index of states} $\boldsymbol{Q}=\{1,\cdots,K\}$
     
\item \textbf{A set of Kraus operators} $ \{ \psi_{x,\mu_x}\}_{x \in \Sigma, \mu_x \in \mathbb{N}} \in \mathbb{C}^{K \times K}$ where  $\sum_{x,\pi} \psi^\dagger_{x,\pi}\psi_{x,\pi}=\mathbb{I}$ and $\dagger$ denotes the complex conjugate transpose.

\item \textbf{An initial state}   $\pi_0 \in \mathbb{C}^{K\times K}$ where {$\pi_0$} is a Hermitian  positive semidefinite matrix of arbitrary rank and $tr(\pi_0)=1$.
\end{enumerate}
Therefore a $K-HQMM$ model is denoted  $qM= (\mathbb{C}^{K \times K}, \{ \psi_{x,\mu_x}\}_{x \in \Theta},\pi_0)$ where $\pi_0 \in \mathbb{C}^{K\times K}$ is the initial state, and $\{ \psi_{x,\mu_x}\}_{x \in \Theta, \mu_x \in \mathbb{N}} \in \mathbb{C}^{K \times K}$  are the Kraus operators. 
In classic context the 3) in  Definition \ref{def:HMM}, we use linear map  $A$ and $B$ to define different actions: emission and transition. In quantum context we use Kraus operators  which characterizes completely any positive map. Moreover by Kraus Theorem any action $\Phi$ to a  quantum state $\rho$ writes 
$\Phi(\rho)= \sum_i \B_i \rho \B_i^\dagger $, with  $\sum_i \B_i^\dagger \B_i =\mathbb{I},$ the complex matrix $\B$ are the Kraus operators. In quantum context both emission and transition are written as action by using Kraus operators.
Therefore the update status $\pi_t$ rule is computed as follows:
 \begin{equation}
     \pi_t=\frac{\sum_{\mu_x} \psi_{x,\mu_x} \pi_{t-1} \psi_{x,\mu_x}^\dagger}{tr(\sum_{\mu_x} \psi_{x,\mu_x} \pi_{t-1} \psi_{x,\mu_x}^\dagger)}
 \end{equation}
 and probability of a given sequence is given by:
 \small
 \begin{equation}
 \PP(X)=tr(\sum_{\mu_{x_t}}\psi_{x_t,\mu_{x_t}},.. (\sum_{\mu_{x_1}}\psi_{x_1,\mu_{x_1}} \pi_0  \psi_{x_1,\mu_{x_1}}^\dagger),.. \psi_{x_t,\mu_{x_t}}^\dagger))
  \end{equation}
\end{definition}

\begin{definition}{The Learning Problem}
\begin{itemize}
    \item \textbf{Given:} The sequence of observations: \\  $X= \{ x_1, \dots, x_L\}$
    \item \textbf{Question:} How do we learn Kraus operators $\{ \psi_{x,\mu} \}$ to model $X$ using an QHMM?
\end{itemize}
Both approaches \cite{srinivasan2018learning,adhikary2020expressiveness} use the negative $\mathcal{L}$ log-likelihood of the data as a loss function. This function can be written as a function of the set of Kraus operators $\{ \psi_{x,\mu} \}$ as follows:
\begingroup
\small
\begin{equation}
\label{eq:Logl}
\mathcal{L}= - \ln \tr(\sum_{\mu_{x_t}}\psi_{x_t,\mu_{x_t}},..(\sum_{\mu_{x_1}}\psi_{x_1,\mu_{x_1}} \pi_0  \psi_{x_1,\mu_{x_1}}^\dagger),.. \psi_{x_t,\mu_{x_t}}^\dagger))
\end{equation}
\endgroup
with the following constraint $\sum_{x,\mu} \psi^\dagger_{x,\mu}\psi_{x,\mu}=\mathbb{I}$.
\end{definition}
An equivalent formulation to the problem of learning a set of $N$ trace-preserving $k \times k$ Kraus operators can be the problem of learning a matrix $\kappaa \in \mathbb{C}^{kN \times k}$ where $N=M\mu$ ($M=|\Sigma|$ and $\mu$ is the number of kraus operators for each observation)  and $\kappaa^\dagger\kappaa=\mathbb{I}$, where the blocks of  $\kappaa$  represent the Kraus operators $\{ \psi_{x,\mu} \}$ that parametrize the QHMM. The initialization of the matrix $\kappaa_0$ is done randomly, provided that it is unitary. The goal is to update this matrix until we find a matrix $\kappaa^*$ that minimizes the function $\mathcal{L}$ in Equation \eqref{eq:Logl}.
\\
\\Srinivasan's approach \cite{srinivasan2018learning} seeks iteratively to find a series of Givens rotations that locally increase the log-likelihood.  But, a Givens rotation only modifies two rows at a time, which leads to the fact that this approach is too slow for learning large matrices. \cite{adhikary2020expressiveness} proposes to solve this problem using gradient descent. It proposes to learn $\kappaa$ directly with the gradient descent method where G denotes the gradient of the loss function $\mathcal{L}$ in Equation \eqref{eq:Logl} where the update function is:
\begin{equation}
\label{eq:upf}
    \kappaa_{i+1}=\kappaa_i-\tau \mathbb{U}_i(\mathbb{I}+\frac{\tau}{2} \V_i^\dagger\U_i)^{-1} \V_i^\dagger\kappaa_i
\end{equation}
where $\U_i =[ G|\kappaa_i]$, $\V_i=[\kappaa_i|-\G]$, and $\G$ is the gradient at $\kappaa_i$. Algorithm \ref{al:HQMM} summarizes the approach performed by \cite{adhikary2020expressiveness} to learn the parameters of QHMMs using the gradient descent. 
\\

\begin{algorithm}[h]
    \SetKwInOut{Input}{Input}
    \SetKwInOut{Output}{Output}
    \SetKwInOut{Return}{Return}

    \Input{Training dataset $X \in \mathbb{N}^{m \times l}$, learning rate $\tau$, learning rate decay $\alpha$, number of batchs $\beta$, number of epochs $\upsilon$.}
    \Output{$\{ \psi_{i} \}_{i=1}^{M\mu}$}
    
    \textbf{Initialisation}
  
Complex orthonormal matrix on Stiefel manifold $\kappaa \in \C^{M \mu n \times n} $ and partition into Kraus operators $\{ \psi_{i} \}_{i=1}^{M\mu}$ whith $ \psi_{i} \in \C^{n \times n}$

    \For{epoch = 1:$\upsilon$}{
    Partition training data $X$ into $\beta$ batches ${\X_\beta}$

        \For{b=1:$\beta$}{
        Compute gradient $\G_i \leftarrow \frac{\partial l}{ \partial \kappaa_i}$ for batch $\{\X_\beta\}$ and loss function $l$.
\\ Compute $\frac{\partial l}{\partial \kappaa}= \G \leftarrow [\G_1, \dots, \G_M\mu]^T$
\\ Construct $\U \leftarrow [\G|\psi]$,$\V \leftarrow [\psi|-\G]$
\\ Update $\kappaa \leftarrow \kappaa-\tau \U(\I+\frac{\tau}{2} \V^\dagger\U)^{-1} \V^\dagger\kappaa$
        
        }
        Update learning rate $\tau = \alpha \tau$
Re-partition $\kappaa$ into $\{\kappaa_i\}$
    }\Return{$\{\kappaa_i\}$} 
    \caption{ Learning QHMMs \cite{adhikary2020expressiveness}}
    \label{al:HQMM}
\end{algorithm}

\section{Highlights of our contribution}
\label{sq:OC}
\noindent In this section, we will give and describe the highlights of our contribution for the problem of searching failure scenarios of a general system noted $S$. To do this, we start by defining what is a system and what is a severe state of a system and also what is a Failure Scenario? 

\begin{definition}{System $S=(\Xi ,\Phi,\Lambda)$: } is a set of $n$ basic events $\Xi =\{\xi_1, \dots, \xi_n\}$  connected between them in a predefined way, where each $\xi_i$ has a probability of breaking down $\mathbb{P}_{down}(\xi_i)$ and a probability of being repaired $\mathbb{P}_{repair}(\xi_i)$. With $\Phi$ it is the set of the Sever States of the system and $\Lambda$ is a set of failure scenarios of the system. 
\end{definition}

\begin{definition}{Sever state $\phi_i \in \Phi$:  } The state where the whole system is considered to be broken, where $\phi_i$ is a subset of the set $\Xi$ where all events $\xi_j \in \phi_i$ are broken. 
\end{definition}

\begin{definition}{Failure Scenario $\lambda_i \in \Lambda$: } is a succession of failures and repair of basic events of the system, where each sequence has a probability calculated like this $\PP(\lambda_i)=\prod_j \PP_{down / repair}(\xi_j)$. 
\end{definition}
\noindent Suppose that we have a system $S=(\Xi ,\Phi,\Lambda)$ of $|\Xi|=n$ basic events, and we search for the failure scenarios of this system $\Lambda$ that have the probability greater than a fixed probability $\PP_{min}$ $(\PP(\lambda_i) > \PP_{min})$ to reach a serious failure state $\phi_i \in \Phi$.
\\In order to answer this question, we represent the states of the system by a state graph, and we search in this graph for this subset of path $\Lambda$ between the current state $\phi_t$ of the system and all the states of the serious failures $\phi_i \in \Phi$.
If we have a system with $n$ basic events, automatically we will have $2^n$ states of the system, which gives a graph of states with $2^n$ vertices. The problem here is that the complexity of finding $k$ paths between two vertices of this graph is NP-complete \cite{al2021complexity}. So, for each element of $\Phi$, it takes NP-complete complexity to find only $k$ paths instead of all paths. Also, if the initial state $\phi_t$ is changed, all calculations must be performed again. This approach becomes impossible if we have a system with a very large number of basic events, which leads us to look for other methods to find these failure scenarios. 
\\Instead of using this approach, we propose to learn QHMMs to represent the set of failure scenarios $\Lambda$ of the system, which allows us to handle these scenarios in a better way, and also to create some predictive models that can be exploited  dynamically to generate different possible scenarios from a given state of the system.  This approach gives us the possibility to process the failure scenarios to all severe failures instead of processing them one-by-one, and it also gives us the possibility to process only the possible scenarios with a high probability.  In addition, this approach gives us the possibility to see if a failure scenario is probable or not. 
\vspace{-.1cm}
\section{Experimental validation}
\label{sq:results}
 
\noindent In this section, we will show the results of the test on a real small systems. Firstly, we start by comparing the results of QHMMs and HMMs using as metric DA described in the following.
\vspace{-.1cm}
\subsection{Metric}
\noindent Description accuracy is a scaled
log-likelihood independent of sequence of length $L$ \cite{zhao2010norm},\cite{srinivasan2018learning}. 
Consider a non linear function 
$f :(-\infty, 1] \rightarrow (-1,1],$ the metric is defined as the following:
\begin{equation}
    DA= f(1+\frac{\log_s \PP(Y/D)}{L})
\end{equation}
\vspace{-.4cm}
where
$$
f(x) =
\left\{
	\begin{array}{ll}
		x,  & \mbox{if } x \geq 0 \\
		\frac{1-\exp{(-x/4)}}{1+\exp{(-x/4)}}, & \mbox{if } x < 0.
	\end{array}
\right. 
$$
Where $L$ is the length of the sequence, $s$ is the number of output symbols in the sequence, $Y$ is the data, and $\mathbb{D}$ is the model. 
When $DA=1$, the model predicted the sequence with perfect accuracy, and when $DA>0$, the model performed better than random. On the real-world dataset, we report the average accuracy for a classification problem.
\subsection{Complexity}
\noindent The complexity of computing the loss $ \mathcal{L}$ using the Equation \eqref{eq:Logl} is 
$O( \mu M L  K^3)$, with $M$ the number of sequences in the batch, $L$  the length of sequences and for the complexity of the update function \eqref{eq:upf} is $O( \mu M  K^3)$. On the other hand, in the classic case, we need to do $K$ operations for each cell which gives the complexity analysis $O(M L K^2)$ for the Forward and Backward algorithms, and we need to store a matrix of size  $K \times  M$ for each cell which gives $O(M L K)$ for space complexity.

\subsection{Test results}
\begin{figure}[ht]
\centering
\includegraphics[scale=0.3]{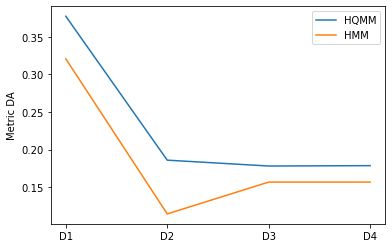}
\includegraphics[scale=0.3]{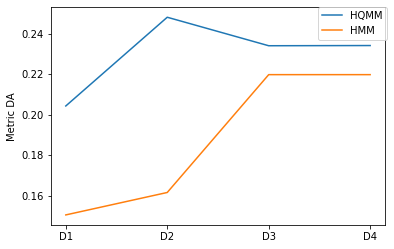}
\caption{ A comparison between HMMs and QHMMs according to the average of DA metric. Figure at left for training dataset and at right for test dataset.  }
\label{resultcom}
\end{figure}

\noindent In this section, we will show the results of the tests on four datasets $D1,D2,D3$ and $D4$ of two small PSA systems $S_1$ and $S_2$. Where, D1 and D2 is the dataset of probable and no-probable scenarios of $S_1$ respectively, D3 and D4 is the dataset of probable and no-probable scenarios of $S_2$ respectively.
\\Firstly, we compare the classical HMMs and the quantum QHMMs methods by using DA metric. For this purpose we compute the average of DA metric in the results obtained by training and tests for the four datasets $D1,D2,D3,D4$ and both approaches HMMs and QHMMs, the results are shown in figure \ref{resultcom}.

\begin{figure}[ht]
\centering
\includegraphics[scale=0.38]{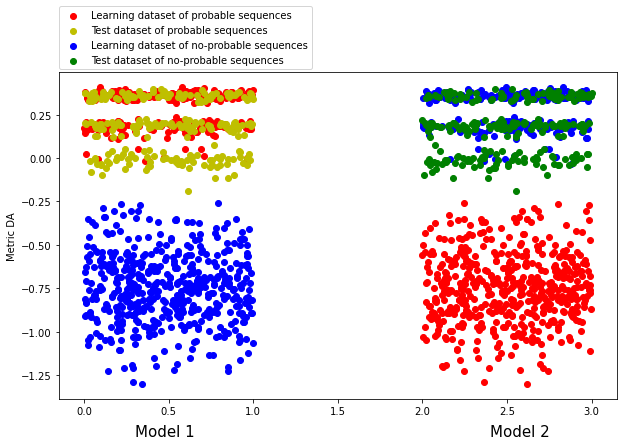} 
\caption{ Results of the two models of the system  $S_1$. }
\label{resultD1}
\end{figure}

\begin{figure}[ht]
\centering
\includegraphics[scale=0.38]{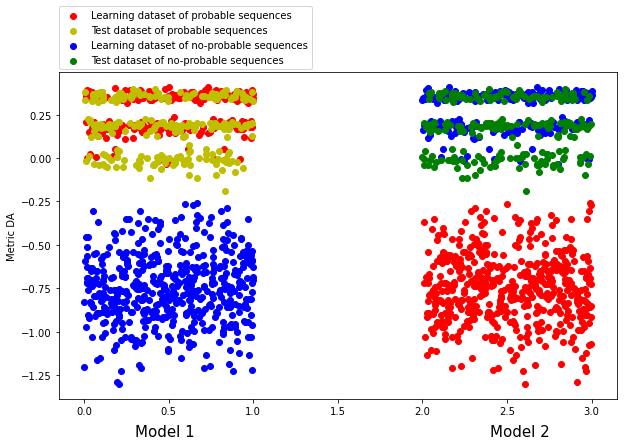} 
\caption{Results of the two models of the system  $S_2$. }
\label{resultD2}
\end{figure}

\noindent From the results in \ref{resultcom}, we can see that the average of DA metric for QHMMs is always higher than HMMs which means that QHMMs is more efficient than HMMs. 
\\In order to identify the probable and no-probable scenarios of a system we use the following strategy: For each system we learn two QHMMs, the first to identify the probable scenarios, and the second to identify the no-probable scenarios. When we have a new scenario we use these two models to decide if it is probable or not, we choose the model that gives the greatest metric DA. In Figure \ref{resultD1}, we show the results of the two models of the first system $S_1$ where blue color represents the no-probable failure scenarios dataset, red color for the training dataset of the probable scenarios, young for the test dataset of the probable scenarios and green color for the test dataset of the no-probable scenarios. The same thing for the second system $S_2$, we show the results in Figure \ref{resultD2}. 

\noindent In these results, we can see clearly that Model 1 and Model 3 are able to detect the probable scenarios and Model 2 and Model 4 are able to detect the no-probable scenarios for the both systems. After that we use those two models of each system in order to get the probable or not probable scenario, to do that we calculate the metric of that sequence for each model and we decide that the sequence is probable or not according to the biggest metric found.  In addition, we use these two models of the system to directly generate the probable and no-probable scenarios from a given state of the system, instead of searching for new ones. To do this, we simply give the current state of the system as input to the two models and generate the requested scenarios. 
\section{Conclusion}
\noindent In the last few years, several quantum algorithms have been proposed for solving various problems in the field of Machine Learning, both in terms of quality and performance as well as in terms of complexity and execution time. In this paper, we have proposed a strategy to learn failure scenarios for PSA system. We have proposed to use QHMMs models to generate failure scenarios from a given state of the system and also to identify the probable and no-probable failure scenarios of a system. This strategy gives us several advantages compared to the current approaches used in the field of PSA, among them, by doing the learning only once and we can find the failure scenarios from any state of the system. In addition, it allows us to generate new scenarios that we do not have in the dataset. To test this approach, we used four datasets for two small real systems to show that the QHMMs is more efficient than the HMMs. Further more, to show that the models are able to detect the probable and no-probable failure scenarios.

\medskip
\bibliographystyle{plain}
\bibliography{references}

\end{document}